\documentclass[%
 reprint,
 amsmath,amssymb,
 aps,
]{revtex4-1}

\usepackage{graphicx}
\usepackage{dcolumn}
\usepackage{bm}

\begin{document}


\title{The evolution of sleep is inevitable}

\author{Jared M. Field}
\email{jared.field AT maths.ox.ac.uk}
\affiliation{%
Wolfson Centre for Mathematical Biology, Mathematical Institute, University of Oxford, Oxford OX2 6GG,
United Kingdom \\
 }%
\altaffiliation[Also at ]{Mathematical Ecology Research Group, Department of Zoology, University of Oxford, Oxford, UK}



\author{Michael B. Bonsall}
\affiliation{
 Mathematical Ecology Research Group, Department of Zoology, University of Oxford, Oxford, UK
}%


\date{\today}

\begin{abstract}
There are two contrasting explanations of sleep: as a proximate, essential physiological function or as an adaptive state of inactivity and these hypotheses remain widely debated. To investigate the adaptive significance of sleep, we develop an evolutionary argument formulated as a tractable partial differential equation model. We allow demographic parameters such as birth and mortality rates to vary through time in both safe and vulnerable sleeping environments. From this model we analytically calculate population growth rate (fitness) for sleeping and non-sleeping strategies. We find that, in a temporally heterogeneous environment, sleeping always achieves a higher fitness than not sleeping. As organisms do not exist in constant environments, we conclude that the evolution of sleep is inevitable. 


\end{abstract}

\maketitle

Most attempts to explain the evolution of sleep, a vulnerable state observed across diverse taxa,  have thus far focussed on a search for benefits of physiological or vital functions. Sleep, it has been proposed, evolved because there is a universal core function that cannot occur during wakefulness. Indeed, it has been suggested that sleep can reduce oxidative stress accumulated during wakefulness \cite{reimund1994free,eiland2002increases, ramanathan2002sleep},  implicated in memory consolidation \cite{mednick2003sleep} and hypothesised to be necessary in heat regulation \cite{mcginty1990keeping}. These and other theories, however, fail to explain either the diversity of sleep patterns observed in nature \cite{campbell1984animal} or the existence of sleep or sleep-like states in some organisms. It is difficult, for example, for the information processing theory of sleep \cite{tononi2006sleep} to account for lethargus, the sleep-like state in \emph{C. elegans} \cite{raizen2008lethargus}.

An alternative perspective is to view sleep as an adaptive state of inactivity \cite{siegel2009}. Sleep and sleep-like states have value insofar as they allow for efficient use of finite energy. Moreover, they may in some instances actually reduce the risk of injury and/or predation \cite{siegel2009}. It is important to note that this adaptive view of sleep evolution does not exclude the existence of other vital functions. Indeed, there is no reason to suspect these functions did not evolve later via exaptation.  A major objection to this view, however,  is that if sleep is adaptive why do we not find organisms that have adapted to not sleep \cite{cirelli2008}? To be more precise, one should expect the costs and benefits of sleep to vary drastically across species in different ecologies and thus, one might expect to find scenarios where the costs outweigh the benefits. Even this question, of whether all organisms have sleep or sleep-like states, is still widely debated \cite{siegel2008}. 

In this paper, we present a model of the adaptive theory of sleep. Our model is formulated as a continuous partial differential equation (PDE) akin to the McKendrick-von F\"{o}rster equation of classical demography \cite{Keyfitz97}.  However,  we allow basal demographic and ecological parameters such as birth and mortality rates to vary through time. We then weight these basal parameters by sleep strategies or functions that quantify activity to find effective mortality and birth rates for a particular strategy. A constant-value sleep function is taken to describe no sleep, whereas oscillations about this value are taken to reflect a sleeping strategy; the cost of higher activity at some times is offset by lower activity (sleep) at other times. We ensure, however, that over a sleep cycle the total amount of activity (defined as the integral over the sleep function) is the same. This way, when we compare the fitnesses of a sleeping and non-sleeping strategy, we are able to evaluate comparable activity strategies.

Our analyses show that when birth or mortality rates are non-constant, there is always a sleep strategy that achieves a fitness higher than the no sleep strategy. Indeed, we show that in a heterogeneous environment the evolution of sleep is inevitable. Further, contrary to the major objection of the adaptive view of sleep, one should only expect to find an organism that does not sleep in a purely constant environment. That is, in the wild we should in fact not expect to find organisms that have evolved to forgo sleep. 

The rest of this paper is organised as follows: In the next section we explain the model and assumptions. Following this, we present analytic calculations of population growth rate, which we use as a measure of fitness \cite{fisher1930genetical, keyfitz2005applied}. We then go on to compare sleeping and non-sleeping strategies under constant and non-constant mortality and birth rates. We additionally consider different sleeping environments where sleeping is assumed to either increase or decrease mortality. Finally, we summarise our findings and suggest future work. 
 
\section*{Model}
We start by defining $n(t,\tau, x)$ the population density of organisms employing a given strategy at time $t$, time they last consumed food $\tau$ and age $x$.  We say that a population has a different strategy if they have a different sleep function $s(t)$ that quantifies how active an organism is at any time $t$. We take large values of $s(t)$ to represent highly active periods and values approaching zero to represent inactive periods. In this way, if we let $\alpha(t)$ be the baseline foraging success rate then $\alpha(t)s(t)$ will quantify the foraging success rate for a strategy with sleep function $s(t)$. With this form, an organism that is active (high $s$) when the resources or prey it consumes are plentiful (high $\alpha$) will be more successful than one that is inactive (low $s$) at the same time.
  
Similarly,  we denote the basal mortality rate $\gamma (t) $ which we weight more generally by a function $f(s)$, giving an effective mortality of $\gamma(t) f(s)$. The functional form of $f$ will change depending on the particular ecological scenario under consideration. In particular, this will alter depending on if we assume sleeping increases or decreases predation rates.  

The rate of change of the population density with sleep strategy $s$ can now be written as 
\begin{equation}
\frac{dn}{dt} = -\gamma(t)f(s)n - \alpha(t)s(t)n. \label{1.1}
\end{equation}  
Successful foraging appears to reduce the population here because if resources are found, the time they last consumed food $\tau$ is reset to zero. In other words, they are not lost but are transferred to the boundary so that when $\tau = 0$ we have 
\begin{equation}
n(t, 0, x) = \alpha(t)s(t) \int_0^d n(t,\tau, x) d\tau, \label{1.2}
\end{equation} 
where $d$ is the maximum time an organism can live without food. 

If we denote the birth rate $\beta(t)$, then similarly when $x=0$ we have
\begin{equation}
n(t, \tau, 0) = \beta(t)s(t) \int_0^m n(t,\tau, x) dx,\label{1.3}
\end{equation}
where $m$ is the maximum life span. In \eqref{1.3}, we assume that activity levels affect effective birth rates as they do foraging successes -- an organism that is sleeping while potential mates are available will enjoy less success than those that are awake. 

Using the chain rule on the right hand side of \eqref{1.1} this becomes
\begin{equation}
\frac{\partial n}{\partial t} + \frac{\partial n}{\partial \tau} + \frac{\partial n}{\partial x} = -\gamma(t)f(s) n - \alpha(t)s(t)n, \label{1.4}
\end{equation}
where, being in the same units, we have taken $d\tau /dt = 1$ and $dx/dt = 1$.

Finally, we close this hyperbolic system \eqref{1.2}-\eqref{1.4} with the initial condition
\begin{equation}
n(0, \tau , x) = n_0(\tau, x). \label{1.5}
\end{equation}

\section*{Population Growth Rate}
To compare different strategies we will eventually use population growth rate as a measure of fitness. To get us there, we start by integrating the left hand side of \eqref{1.4} with respect to $\tau$. Doing so gives
\begin{multline}
\int _0^d \left(\frac{\partial n}{\partial t} + \frac{\partial n}{\partial \tau} + \frac{\partial n}{\partial x}\right) d\tau = \\ \int ^d_0 \frac{\partial n}{\partial t} d\tau + n(t, d, x) - n(t, 0, x) + \int ^d_0 \frac{\partial n}{\partial x} d\tau ,  \label{2.1}
\end{multline}
by the Fundamental Theorem of Calculus. As $d$ is the maximum time an organism can survive without food, such that $n(t, d, x) = 0$ and $n(t,0,x)$ is given by \eqref{1.2} we can write the right hand side of \eqref{2.1} as 
\begin{equation} 
 \int ^d_0 \frac{\partial n}{\partial t} d\tau  - \alpha(t)s(t) \int_0^d n(t,\tau, x) d\tau + \int ^d_0 \frac{\partial n}{\partial x} d\tau. \label{2.2}
\end{equation}
If we define $N(t)$ such that 
\begin{equation}
N(t) = \int^m_0\int^d_0 n(t,\tau, x)d\tau dx,
\end{equation} 
which is the total population at any time and integrate the left hand side of \eqref{2.1} and \eqref{2.2} with respect to $x$ we get
\begin{multline}
\int^m_0\int _0^d \left(\frac{\partial n}{\partial t} + \frac{\partial n}{\partial \tau} + \frac{\partial n}{\partial x}\right) d\tau dx = \\
\frac{dN}{dt}  - \alpha(t)s(t)N(t) + \int ^d_0 n(t, \tau, m) - n(t, \tau, 0) d\tau ,\label{2.4}
\end{multline}
again by the Fundamental Theorem of Calculus and by assuming $n$ is sufficiently smooth so we can change the order of integration. As $m$ is the  maximum life span $n(t, \tau, m) = 0$ and as $n(t, \tau, 0)$ is given by \eqref{1.3} we can write \eqref{2.4} as
\begin{multline}
\int^m_0\int _0^d \left(\frac{\partial n}{\partial t} + \frac{\partial n}{\partial \tau} + \frac{\partial n}{\partial x}\right) d\tau dx = \\
\frac{dN}{dt}  - \alpha(t)s(t)N(t) -\beta(t)s(t) N(t) .\label{2.5}
\end{multline}

Finally, performing the same integrations, but on the right hand side of \eqref{1.4}, and equating to the right hand side of \eqref{2.5}, we find that the dynamics of $N$ are described by 
\begin{equation}
\frac{dN}{dt} = \left(\beta(t)s(t) - \gamma(t)f(s) \right) N,
\end{equation}
which has solution 
\begin{equation}
N(t) = N(0)e^{\int^t_0 \beta(\rho)s(\rho)d\rho - \gamma(\rho)f(s(\rho)) d\rho}.
\end{equation}
Hence the growth of any population with strategy $s(t)$ will be characterised by  
\begin{equation}
r= \int^t_0 \left(\beta(\rho)s(\rho) - \gamma(\rho)f(s(\rho))\right) d\rho. \label{2r}
\end{equation} 
 
\section*{Constant Birth \& Mortality Rate}
For the moment, we assume that $\gamma(\rho) = \Gamma$ is constant and $f(s)=1$ so that mortality is unaffected by sleep.  We start by comparing the fitness of organisms with different sleep functions when the birth rate has the constant value $\beta(\rho) = B$. We denote the fitness of an organism with constant activity ($s(\rho) = 1$) by $r_1$ which, from \eqref{2r}, is given by
\begin{equation}
r_1 = Bt - \Gamma t.
\end{equation}

\begin{figure}

\caption{Sleep function $s(\rho) = 1+ \cos(\rho)$  oscillating about the constant case of $s(\rho) = 1$. Observe the cost of higher activity at some times is lower activity at others.
}

\includegraphics[scale=0.6]{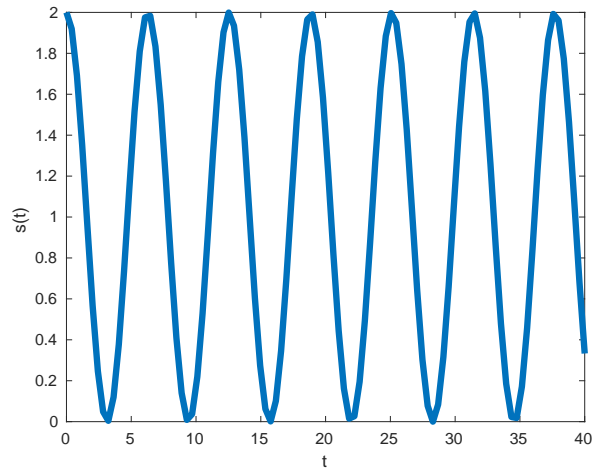}
\label{fig:Fig1}
\end{figure}

We take the simple function $s(\rho) = 1 + \cos\rho$ as an example sleep function. This way, the activity of this phenotype oscillates about the awake case of $s(\rho) = 1$, benefiting from higher activity levels at some times at the cost of lower activity at others (as in Fig. \ref{fig:Fig1}). We do not pretend that this form will coincide with the sleep pattern of any organism in particular. It is a convenient fiction that aids the demonstration of a principle. We denote the fitness of this phenotype as $r_2$ which, again by \eqref{2r}, is given by
\begin{equation}
r_2 = Bt - \Gamma t + B\sin t.
\end{equation}
Note that $r_2$ oscillates about $r_1$ so that on average neither phenotype will have a higher fitness than the other. In other words, sleeping is selectively neutral. 

\section*{Variable Birth Rate \& Constant Mortality Rate}
We now consider a birth rate that oscillates about the constant case. This variation may arise for a variety of reason, and may include availability of resources or availability of mates. In particular, we take $\beta (\rho) = B\left(1+ \cos \rho \right)$. In this case, we find $r_1$ to be given by 
\begin{equation}r_1 = Bt - \Gamma t + B\sin t,\end{equation}
whereas $r_2$ is given by 
\begin{equation}r_2 =\frac{3Bt}{2} - \Gamma t + 2B\sin t + \frac{B}{4}\sin 2t .\end{equation}
Oscillations aside, observe that the coefficient of $t$ is larger in $r_2$ than in $r_1$ so that for almost all $t$
\begin{equation}
r_2 > r_1.
\end{equation}
\begin{figure}

\caption{Typical population growth rates $r_1$ (dotted line) and $r_2$ (solid line) as a function of time when birth rates are variable and mortality is constant. Observe that $r_2 > r_1$ for almost all $t$. Here $B=5$ and $\Gamma = 2.5$.
}

\includegraphics[scale=0.6]{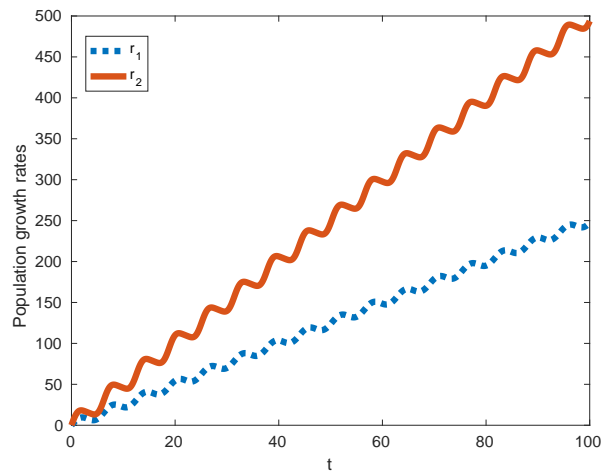}
\label{fig:Fig2}
\end{figure} 
In Fig. \ref{fig:Fig2} we present a typical plot of $r_1$ (dashed line) and $r_2$ (solid line) as a function of time, showing that $r_2 > r_1$. The precise parameter values have no significance of themselves but demonstrate that for large times the oscillations are unimportant.  

Hence, subject to a constant mortality rate and oscillating birth rate one should expect an organism that sleeps to have a greater fitness than one that does not. 

\section*{Constant Birth Rate \& Variable Mortality Rates}
In the case where birth rate and activity are constant but mortality is variable such that $\gamma (\rho) = \Gamma (1+\cos \rho)$, again from \eqref{2r}, we find that 
\begin{equation}
r_1 = Bt -\Gamma t - \Gamma \sin t 
.\end{equation}
\subsection*{Sleep in a Vulnerable Environment}
We now consider the case where $f(s) = \max s -s$, is a decreasing function of $s$. This way, if an organism is sleeping (low $s$) it increases its mortality rate. This we take to model the situation of an organism sleeping in an open environment or non-socially so that the vulnerability associated with sleep increases predation. So in the case of our simple sleep function we have $f(s) = 1- \cos \rho$. We continue to assume that birth rates are unaffected by sleep. In this instance, we denote the associated fitness by $r_{2v}$ which, by \eqref{2r}, takes the value
\begin{equation}
r_{2v} = Bt - \frac{\Gamma}{2}t + \frac{\Gamma}{4}\sin 2t.
\end{equation}

\subsection*{Sleep in a Safe Environment}
\begin{figure}

\caption{Typical population growth rates $r_1$ (dotted line),  $r_{2s}$ (solid line) and $r_{2v}$ (coincides with $r_{2s}$) as a function of time when birth rates are constant and mortality is variable. Observe that $r_{2s} = r_{2v} > r_1$ for almost all $t$. Here $B=5$ and $\Gamma = 2.5$.
}

\includegraphics[scale=0.6]{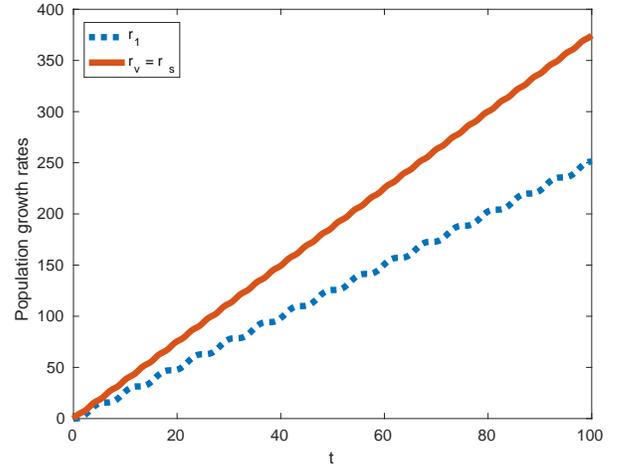}
\label{fig:Fig3}
\end{figure}

In a safe environment whereby sleeping would be expected to decrease predation, we take $f(s) = s$. In this case, we choose the sleep function such that $s(\rho) = 1- \cos \rho$. We let the fitness under these conditions be given by $r_{2s}$, which is found to be
\begin{equation}
r_{2s} = Bt - \frac{\Gamma}{2}t + \frac{\Gamma}{4}\sin 2t.
\end{equation}
Clearly, for almost all $t$ we then have the following inequalities:
\begin{equation}
 r_{2v}> r_1,
\end{equation}
\begin{equation}
r_{2s} > r_1.
\end{equation}
In other words, when birth rates are constant but mortality rates oscillate there exists a sleep function in both safe and vulnerable environments such that an organism that sleeps enjoys a higher fitness than one that does not (again see Fig. \ref{fig:Fig3} for a typical example). Intuitively, in the vulnerable case it is best to stay most active during periods with the highest mortality. Whereas in the safe environment it is more beneficial to shift activity such that the peaks occur when mortality is lowest.

\section*{Variable Birth \& Mortality Rates}
We now consider the most general case where both birth and mortality are non-constant and affected by activity. We take the mortality function as before so that $\gamma( \rho) = \Gamma \left(1 + \cos \rho \right) .$ As there is no particular reason to assume that birth rates will be in phase with mortality rates, we now take $\beta(\rho) = B \left(1+ \cos(\rho-g)\right)$, for a constant $g \in [0, 2\pi)$. Again by \eqref{2r}, we find for the awake strategy $s(\rho) = 1$, the fitness given by
\begin{equation}
r_1 = \left(B-\Gamma\right)t + B\cos g \sin t - B\sin g\cos t - \Gamma \sin t + B\sin g. 
\end{equation}

\subsection*{Sleep in a Safe Environment II}
Recall that in a safe environment $f(s) = s$, so that sleeping reduces mortality. In this case, we take the general form $s(\rho) = 1+ \cos (\rho - q)$, where $q \in [0, 2\pi)$ is a constant phase shift. This way, we can quantify the fitness of a sleep strategy that is possibly out of phase by $q$, in an environment where the birth rate is possibly out of phase by $g$. This time we present only the part of the fitness that is not oscillatory, which for long times is sufficient to compare fitnesses. We do however, present the full value and details of the calculation in the Supporting Information. We find the non-oscillating fitness under the above conditions to be given by 
\begin{equation}
r_{2s} = \left(B-\Gamma\right) t + \frac{Bt}{2}\sin q \sin g + \frac{Bt}{2}\cos g\cos q - \frac{\Gamma t}{2}\cos q. \label{2s}
\end{equation}

\subsection*{Sleep in a Vulnerable Environment II}
As before, in a vulnerable environment we take $f(s) = \max s - s$. Hence, for the general sleep function $s(\rho) = 1+ \cos (\rho - w)$ we find that $f(s) = 1- \cos (\rho - w)$, where $w$ is another constant phase shift. Here we also only present the non-oscillatory part of the fitness, which is given by
\begin{equation}
r_{2v} = \left(B-\Gamma\right) t + \frac{Bt}{2}\sin w \sin g + \frac{Bt}{2}\cos g\cos w + \frac{\Gamma t}{2}\cos w. \label{2v}
\end{equation}

\begin{figure}

\caption{Typical population growth rates $r_1$ (dotted line),  $r_{2s}$ (solid line) and $r_{2v}$ (coincides with $r_{2s}$) as a function of time when birth rates and mortality rates are variable. Observe that $r_{2s} = r_{2v} > r_1$ for almost all $t$. Here, $B=5$ , $\Gamma = 2.5$ and $g= \pi/3$. 
}

\includegraphics[scale=0.6]{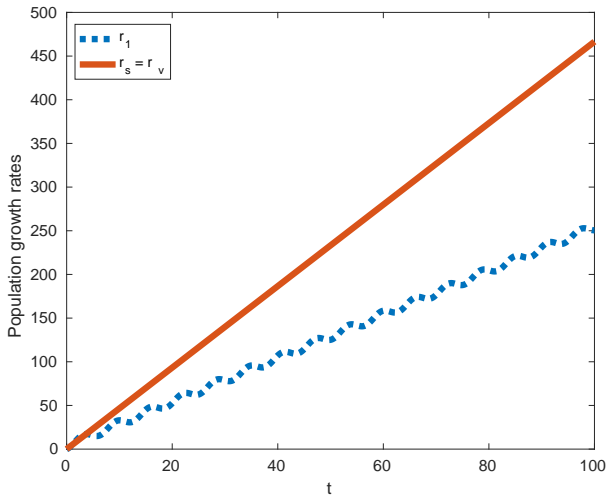}
\label{fig:Fig4}
\end{figure}

Observe that the non-oscillatory part of $r_1$ is given by $\left(B - \Gamma \right) t$, which appears in both $r_{2s}$ and $r_{2v}$. Notice that if for any given $g$ we pick 
\begin{equation}
q = \begin{cases}
\frac{\pi}{2} & \text{if} \quad \sin g \ge 0, \\
\frac{3\pi}{2}& \text{if} \quad \sin g < 0, 
\end{cases}
\end{equation} 
then the extra terms in \eqref{2s} are always postive. Similarly in \eqref{2v}, the same is true if for any given $g$ we pick 
\begin{equation}
w = \begin{cases}
\frac{\pi}{2} & \text{if} \quad \sin g \ge 0, \\
\frac{3\pi}{2}& \text{if} \quad \sin g < 0 .
\end{cases}
\end{equation}
While these sleeping strategies may not be the most optimal, we have nonetheless shown that in both environment types, with any degree of asynchrony between birth and mortality rates, there always exists a sleeping strategy that enjoys a higher fitness than constant activity. As before, we provide a typical example of this case in Fig. \ref{fig:Fig4}.

\section*{Discussion}
Here we developed a tractable model to investigate when, if at all, sleep may be adaptive. This model allowed demographic parameters such as birth rate and mortality rate to (potentially) oscillate in time about basal values. We then defined individual sleep strategies or functions, $s(t)$, that quantified the activity of an organism through time. We took $s(t) = 1$ to model an organism that remains awake indefinitely, whereas oscillations about this value modeled sleep. The cost of higher activity at some times was lower activity (sleep) at other times. These functions were then used as weights to find effective demographic values for an organism employing a given strategy. With this set-up, we were then able to compare the fitness (defined as population growth rate) of sleeping and non-sleeping strategies under an array of conditions.

When birth rates were allowed to vary but mortality kept constant, we found that a sleeping strategy achieved a higher fitness than remaining active indefinitely. We then kept birth rates constant and instead allowed mortality to vary. This split into the two cases of when sleeping would increase or decrease mortality. In both instances however, the sleeping strategy had a higher fitness. Intuitively, in a safe sleeping environment it was best to be most active when mortality was lowest. Whereas  in a vulnerable sleeping environment, the converse was found to be true. 

Both mortality rates and birth rates were then allowed to vary at the same time, potentially out of phase. Yet again, we found that in both environment types, there is always a sleep strategy that trumps staying awake indefinitely. Note that the sleep strategies we found in this case were not necessarily the most optimal. However, if there are strategies that are more optimal they must be of the sleeping type. 

The only instance where constant activity has a fitness as good as sleeping was found to be when birth and mortality rates are constant. However, organisms do not exist in a constant world. This result nonetheless highlights that the adaptive theory of sleep is testable.  Indeed, in a recent study on \emph{Drosophila}, 
sleep duration was observed to change adaptively in response to environmental change \cite{slocumb2015enhanced}. Amongst others, this suggests that model organisms such as \emph{Drosophila} have the potential for testing evolutionary and ecological theories of sleep. Designing experiments where demographic variablity can be controlled and hence  environmental constancy might be approximated quite well is a challenge for future work.

If the adaptive value of sleep relates to the efficient use of energy in variable environments, then  why not simply evolve a state of rest? As in \cite{siegel2009}, we suggest that sleep and other quiescent states are in fact best viewed on a continuum. For instance, there is growing evidence that the dormancy in animals and plants evolves in response to varying environmental cues. A recent study argues that seed dormancy emerged at the inception of seed plants due to environmental variability \cite{willis2014evolution}. While in animals, such as mosquitoes, dormancy and diapause are intimately associated with critical photoperiod length and latitudinal variation \cite{bradshaw1977evolution}. However, in general, the evolution of these periods of inactivity are best broadly viewed as adaptations towards the evolution of risk-averse strategies in fluctuating environments \cite{venable1988selective}.

One caveat to our findings might be that our approach lacks the specific details to show the inevitable evolution of sleep, \emph{per se}. While we acknowledge that the physiological characteristics of sleep are complex (REM and non-REM waves in some but not all organisms \cite{siegel2008}; sleep rebound in some but not all organisms \cite{lyamin1998organization,rattenborg2004migratory, ridgway2009dolphins}) and that these can allow sleep to be distinguished from periods of rest (when necessary), we emphasize that sleep (as a strategy) is best viewed on a continuum. Furthermore, given the varied characterisations of sleep \cite{siegel2008}, our results demonstrate the adaptive value of, arguably, sleep’s most defining feature.

Our analyses have hence shown that the evolution of sleep and sleep-like states is inevitable in variable environments. Sleep as a behaviour is,  in and of itself, valuable. While much research has been done to find vital functions that explain why organisms sleep \cite{campbell1984animal, reimund1994free,eiland2002increases, ramanathan2002sleep, mednick2003sleep}, here we have provided broad ecological reasons applicable to diverse taxa. This is not to say that these vital functions do not exist. Undeniably some of them do. However, they are not initially needed for sleep to evolve. Indeed, our analyses plausibly suggest that sleep first evolved simply because activity-inactivity cycles are adaptive in a non-constant world. 

Given a certain ecological context we showed that there is always a sleeping strategy that gained a higher fitness than not sleeping. All of these strategies however changed only the amount of activity at given times. In reality, organisms change the frequency of their sleeping cycles and the length of inactive periods. In future work, it will be interesting to investigate this diversity. In particular, can we specify demographic and ecological parameters and \emph{generate} optimal sleep patterns for those values? 

\section*{Acknowledgments}
JMF is funded by the Charles Perkins Scholarship with additional financial support from UTS, Sydney. We thank Thomas W. Scott for valuable comments and discussion.


\section*{Supporting Information} 
Here we present the detailed calculation to determine the population growth rate as in \eqref{2s}. While we only present $r$ for the case of a safe environment, with variable birth and mortality rates, the integrals performed here include all of the integrals necessary to calculate every other population growth rate presented. 

As outlined in the main part of the paper to find $r_{2s}$  we need to calculate 
\begin{equation}
r_{2s} =  \int^t_0 \beta(\rho)s(\rho)d\rho - \gamma(\rho)f(s(\rho)) d\rho.
\end{equation} 

We split this larger calculation into the two smaller integrals given by 
\begin{equation}
I_1 =  \int^t_0 \beta(\rho)s(\rho)d\rho,
\end{equation}
\begin{equation}
I_2 = \int^t_0 \gamma(\rho)f(s(\rho)) d\rho.
\end{equation}

In the case we are concerned with we take $\beta(\rho) = B \left(1+ \cos(\rho-g)\right)$ and $s(\rho) = 1+ \cos (\rho - q)$ so that, in fact, 
\begin{equation}
I_1 = \int ^t_0 B\left(1+ \cos(\rho-g)\right)\left(1+ \cos (\rho - q)\right)d\rho,
\end{equation}
expanding this gives
\begin{equation}
I_1 = \int ^t_0 B\left(1+ \cos(\rho-g) + \cos (\rho - q) + \cos(\rho -1)\cos(\rho-q)\right) d\rho,
\end{equation}
Using the standard sum of angles formula $\cos(\rho-g) = \cos(\rho)\cos(g) +\sin(\rho)\sin(g)$, the first three terms are trivial to calculate.

We use the same sum of angles formula to expand the fourth term. Doing so, and collecting terms gives  
\begin{multline}
\cos(\rho-q)\cos(\rho-g) = \\\sin q \sin g + \cos ^2 \rho \cos(g+q) + \cos \rho \sin \rho \sin(g+q),
\end{multline} 
where we have used the sum of angles formula again to get the terms involving $g+q$.

Hence, to calculate $I_1$ we need in fact to calculate \begin{equation}
G_1 = \int ^t_0 \cos^2\rho d\rho,
\end{equation}
\begin{equation}
G_2 = \int ^t_0 \cos\rho \sin \rho d\rho.
\end{equation}
We focus attention first on $G_1$. As $\cos^2\rho = \frac{1}{2}\left( 1 + \cos 2\rho\right)$, from standard double angle formula, we find that
\begin{equation}
G_1 = \frac{1}{2}\left(t+ \frac{1}{2}\sin 2t\right)
\end{equation}
Using integration by parts to calculate $G_2$ we find that 
\begin{equation}
G_2 = \frac{1}{2}\sin^2 t
\end{equation}
Putting this all together we find 
\begin{multline}
I_1 = B(t+ \sin t\left(\cos g + \cos q\right) - \cos t \left(\sin g + \sin q\right) + \sin g + \sin q \\
t\sin q \sin g + \frac{1}{2}\cos (g+ q) \left(t + \frac{1}{2}\sin 2t\right) + \frac{1}{2}\sin (g+ q) \sin^2 t ). \label{ref1}
\end{multline}

Recall that, in a safe sleeping environment we took $f(s) = s$ and $\gamma (\rho ) = \Gamma \left(1+ \cos \rho\right)$. Hence, to calculate $I_2$ we simply need to replace $B$ with $\Gamma$ and set $g=0$ in \eqref{ref1}. It follows then that
\begin{multline}
I_2 = \Gamma(t+ \sin t\left(1 + \cos q\right) - \cos t \sin q + \sin q \\
 + \frac{1}{2}\cos q \left(t + \frac{1}{2}\sin 2t\right) + \frac{1}{2}\sin q \sin^2 t ). \label{ref2}
\end{multline}
Finally, if we take the difference of \eqref{ref1} and \eqref{ref2} the non-oscillatory parts are as in \eqref{2s} in the main text, if the $\frac{Bt}{2}\cos(g+q)$ term is expanded once more.

\bibliographystyle{plain}
\bibliography{ref_2016}
\end{document}